\documentclass{article}
\usepackage{pdfpages}
\usepackage{float}
\usepackage{bbm}
\usepackage{amssymb}
\usepackage{multicol}
\usepackage{amsmath}
\usepackage{authblk}
\usepackage{algorithm}
\usepackage{amsfonts}
\usepackage{graphicx}
\usepackage{booktabs}
\usepackage{caption}
\usepackage{siunitx}
\usepackage{multirow}
\usepackage{helvet}
\usepackage[T1]{fontenc}
\usepackage[colorlinks=true, allcolors=red]{hyperref}

\DeclareUnicodeCharacter{2212}{-}

\usepackage[a4paper,top=2.5cm,bottom=2.5cm,left=1.3cm,right=1.3cm,marginparwidth=1.75cm]{geometry}

\usepackage[
    backend=biber,
    style=ieee,
    natbib=true,
    url=false,
  ]{biblatex}

\AtEveryBibitem{\clearfield{month}}
\AtEveryBibitem{\clearfield{day}}

\title{The brain as a blueprint: a survey of brain-inspired approaches to learning in artificial intelligence}

\author[1]{Guillaume Etter}

\affil[1]{Groningen Institute for Evolutionary Life Sciences, University of Groningen, Nijenborgh 7, 97473
Groningen, the Netherlands}
\affil[]{Correspondance: {g.etter}@rug.nl}
\date{
  \parbox{\linewidth}{\centering%
  \today\endgraf\bigskip}
  }

\begin{document}
\maketitle

\begin{abstract}
Inspired by key neuroscience principles, deep learning has driven exponential breakthroughs in developing functional models of perception and other cognitive processes.
A key to this success has been the implementation of crucial features found in biological neural networks: neurons as units of information transfer, non-linear activation functions that enable general function approximation, and complex architectures vital for attentional processes.
However, standard deep learning models rely on biologically implausible error propagation algorithms and struggle to accumulate knowledge incrementally. 
While, the precise learning rule governing synaptic plasticity in biological systems remains unknown, recent discoveries in neuroscience could fuel further progress in AI.
Here I examine successful implementations of brain-inspired principles in deep learning, current limitations, and promising avenues inspired by recent advances in neuroscience, including error computation, propagation, and integration via synaptic updates in biological neural networks.
\end{abstract}

\begin{multicols}{2}
\raggedcolumns
\section{Introduction}
Biological systems adapt to their environments through evolution.
This form of adaptation occurs randomly, through genetic permutations, and can be found in the simplest unicellular animals.
Lifelong learning, on the other hand, refers to the act of improving behaviors throughout one's life.
This process is thought to be directly related to brain activity and involves updating synapses, the points of contact between neurons.
In humans, learning can lead to fast improvements in motor (e.g. learning to play tennis) or cognitive (e.g. mastering the game of chess, learning a new language, ...) abilities.

\begin{figure}[H]
  \includegraphics[width=\columnwidth]{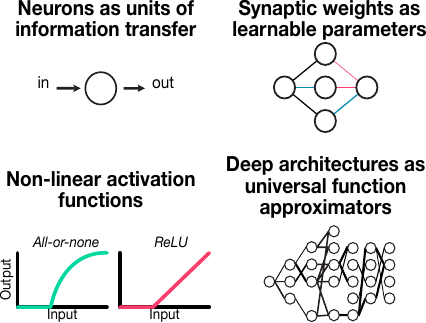}
  %\captionsetup{width=\columnwidth}
  \caption{
    \textbf{Brain-inspired principles of deep learning}. 
    Deep learning implements four key principles from neuroscience: neurons as units of computation, synapses as learnable parameters, non-linear activation functions, and deep architectures that act as universal function approximators.
     }
  \label{fig:fig_basic_principles}
\end{figure}

In recent years, artificial intelligence (AI) has come a long way in replicating learning for these cognitive tasks, including vision \cite{lecunConvolutionalNetworksImages1995, dosovitskiyImageWorth16x162021c}, speech \cite{oordWaveNetGenerativeModel2016}, language \cite{radfordImprovingLanguageUnderstanding}, and problem solving \cite{silverMasteringGameGo2016}.
This success has been fueled by deep learning frameworks \cite{lecunDeepLearning2015}, which implements key elements from biological neural networks: (1) neurons as units of information transfer \cite{mccullochLogicalCalculusIdeas1943, rosenblattPerceptronProbabilisticModel1958a}, (2) nonlinear "all-or-none" \cite{lucasAllNoneContraction1909, adrianImpulsesProducedSensory1926} activation functions \cite{householderTheorySteadystateActivity1941}, (3) deep architectures that can act as universal function approximators, and (4) synaptic weights as learnable parameters (\textbf{Fig.} \ref{fig:fig_basic_principles}).

\begin{figure}[H]
  \includegraphics[width=\columnwidth]{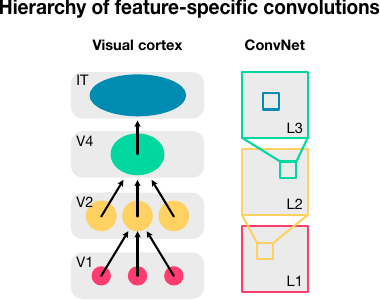}
  %\captionsetup{width=\columnwidth}
  \caption{
    \textbf{Convolutional neural networks take inspiration from the visual system}.
    The visual system is composed of a hierarchy of feature-selective neural networks that can be composed to learn complex visual representations.
    Convolutional neural networks take a direct inspiration from this principle by using convolutions in place of one-to-one connectivity patterns found in multilayer perceptrons.
     }
  \label{fig:fig_convnet}
\end{figure}

It is remarkable that implementing these principles from biological neural networks is sufficient to design models capable of emulating advanced aspects of human cognition.
After all, these models omit much of the complexity observed in the central nervous system, including intricate architectures and various functional cellular types, from inhibitory interneurons to glial cells.
%%% FIGURE
%%% END FIGURE
Beyond the aforementioned basic principles of deep learning, progress in machine learning is still consistently fueled by breakthroughs in neuroscience.
For example, convolutional neural networks were inspired by the architecture of the mammalian visual system \cite{hubelReceptiveFieldsFunctional1968}, specifically by implementing hierarchical feature extraction from input patterns \cite{fukushimaCognitronSelforganizingMultilayered1975, fukushimaNeocognitronHierarchicalNeural1988, fukushimaNeocognitronHandwrittenDigit2003, lecunConvolutionalNetworksImages1995}, leading to visual models that could scale to large and complex images (\textbf{Fig.} \ref{fig:fig_convnet}).
More recently, vision transformers implement attention mechanisms that are essential in human perception \cite{dosovitskiyImageWorth16x162021c}.

%%% FIGURE
%%% END FIGURE

While there is also a long history of unsuccessful attempts at implementing neuroscience principles into artificial models, historically most breakthroughs in AI are somewhat inspired by aspects of the brain.
In parallel, this also suggests that failures to implement aspects of the brain into AI models often reflects limits in our understanding of biological computations.
This prompts a need to regularly evaluate current advances in neuroscience, and translate them into artificial models to not only advance machine learning but also test theories of brain function.

Here, I review recent advances in neuroscience, including in vivo plasticity mechanisms, interneuron microcircuits, oscillations, and the potential role of offline replay in continual learning. Throughout this exploration, I review attempts at implementing these advances in machine learning and what can be learned from the success and failures of these experimentations.
Finally, I briefly cover emerging frameworks including neuromorphic hardware, which has the potential to fuel future implementations of biologically plausible learning algorithms directly into physical hardware.

\section{Towards advanced bioinspired neural networks}
Learning can be broken down into distinct problems including: (1) error computation and propagation, (2) credit assignment, (3) synaptic updates.
This breakdown is mostly useful for reviewing advances in these areas, nevertheless it is noteworthy that these problems are often entangled in practice.
Later, I will also cover the problem of continual learning where the objective is to train neural networks on new tasks continuously without forgetting previous tasks.

\subsection{Computing errors locally with inhibitory interneurons}
Crucially, biological and artificial systems require \emph{some} feedback to know whether behavioral outputs are appropriate for a given task.
This is referred to as the error signal.
If an error signal is zero, it is assumed that the system (model or agent) is performing optimally and no synaptic updates need to be performed.
When nonzero, the error signal provides a feedback for how much worse output behaviors are from some target (or objective).
In machine learning, this objective is often explicit in supervised training contexts, while in biological systems, the objective may only be survival, and error signals could take the form of physiological signaling and/or more specialized signals.
These error signals can be global or local, and are crucial to providing credit assignment during learning.
In the brain, it was proposed that dopamine release encodes temporal difference errors, a form of global error signal \cite{maesCausalEvidenceSupporting2020a}.
Such global error signals are likely not the main drivers of credit assignment \cite{lillicrapBackpropagationBrain2020a}, but could be used in conjunction to feedback signals and local inhibitory drive (see below) to gate and set the sign of synaptic updates \cite{roelfsema_control_2018}.
Beyond dopamine which has been fairly well described in the brain and will not be covered here, local error computations (at the level of a single hierarchical level, or microcircuit) remain elusive.
Here, we can examine the main classes of interneurons in the brain, how they interact within a microcircuit, and how they could contribute to computing local error signals.

A key distinction between artificial and biological neural networks is that in the former, all units are typically comparable in function, whereas in the latter, some neurons release excitatory neurotransmitters while other release inhibitory neurotransmitters.
This dichotomy between excitatory and inhibitory neurons is referred to as 'Dale's principle' \cite{kandelDalesPrincipleFunctional1957}, which ANNs typically do not respect.

The nomenclature of interneurons is complex, typically defined both by function, morphology, and genetic expression, and encompasses hundreds of subtypes \cite{buzsakiHippocampalGABAergicInterneurons2001, markramInterneuronsNeocorticalInhibitory2004a}.
In spite of this broad diversity, main functional classes have emerged in the past few years.

\subsubsection{Parvalbumin interneurons}
First, parvalbumin-expressing (PV) interneurons represent the largest class of cortical inhibitory cells.
They are typically fast-spiking and primarily target the perisomatic region (cell body and proximal dendrites) of principal neurons \cite{huInterneuronsFastspikingParvalbumin2014}.
This strategic targeting allows PV cells to exert powerful control over the output spiking activity of principal, excitatory neurons.
Functionally, PV interneurons are critical for establishing network stability through strong feedback and feedforward inhibition, preventing runaway excitation \cite{pouilleEnforcementTemporalFidelity2001}.
They also play a key role in synchronizing neural populations in theta (around 8 Hz) \cite{amilhonParvalbuminInterneuronsHippocampus2015} and gamma (30 - 120 Hz) \cite{wulffHippocampalThetaRhythm2009}, frequency bands, which is thought to be instrumental for temporal coding, feature binding, and routing information across brain regions \cite{cardinDrivingFastspikingCells2009, buzsakiMechanismsGammaOscillations2012}.
By enforcing precise spike timing and controlling overall network gain, PV interneurons create a stable yet dynamic environment conducive to reliable neural computation.
Importantly, PV neurons also directly gate learning, and their activity decreases upon reaching goals in goal-directed tasks \cite{jeongGoalspecificHippocampalInhibition2025}.

\subsubsection{Somatostatin interneurons}
In contrast to the perisomatic targeting of PV cells, somatostatin-expressing (SOM) interneurons primarily innervate the dendrites of principal neurons, particularly the apical tufts where many excitatory inputs arrive \cite{urban-cieckoSomatostatinexpressingNeuronsCortical2016, watersSupralinearCa2Influx2003, larkumDendriticSpikesApical2007}.
SOM neurons typically exhibit adapting firing patterns and respond well to persistent activity.
Their dendritic inhibition allows them to selectively filter or gate synaptic inputs, influencing dendritic integration and the conditions required for triggering dendritic spikes, which are themselves powerful computational events \cite{larkumSynapticIntegrationTuft2009}.
By controlling dendritic excitability, SOM interneurons can modulate synaptic plasticity occurring at dendritic synapses.
% Add O-LM neurons, martinoti
They essentially act as a gating mechanism on dendritic activity and associated plasticity \cite{tremblayGABAergicInterneuronsNeocortex2016}.
Their potential role in learning is even more striking when considered together with VIP interneurons.

\subsubsection{VIP interneurons}
\begin{figure}[H]
  \includegraphics[scale=1.5]{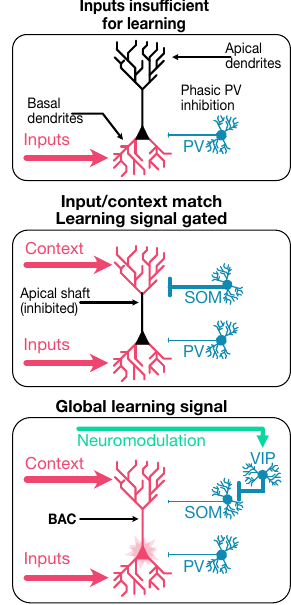}
  %\captionsetup{width=\columnwidth}
  \caption{
    \textbf{Inhibitory microcircuit}. 
    Top, parvalbumin (PV) neurons contribute to maintaining neurons in an equilibrium state, whereby weak activation of basal dendrites by inputs is insufficient to elicit output action potentials.
    Middle, when apical dendrites are activated by contextual inputs, somatostatin (SOM) neurons prevent backpropagation of calcium (BAC) signals by inhibiting the apical shaft.
    Bottom, when global learning signals are provided through neuromodulation, vasoactive intestinal peptide (VIP) provide feed-forward disinhibition of the apical shaft and allow BAC signals to propagate.
     }
  \label{fig:fig_inhibitory_microcircuit}
\end{figure}

Vasoactive Intestinal Peptide-expressing (VIP) interneurons constitute a third major class, unique in that they predominantly target other interneurons, most notably SOM cells \cite{piCorticalInterneuronsThat2013}.
This creates a feedforward disinhibitory circuit: activation of VIP neurons inhibits SOM neurons, which in turn reduces the inhibition onto the dendrites of principal neurons.
This disinhibition effectively opens a gate for dendritic activity and plasticity (which I will cover in more details below).
VIP neurons are heavily modulated by top-down signals, including reinforcement signals and attentional cues, often conveyed by neuromodulators like acetylcholine and norepinephrine \cite{letzkusDisinhibitoryMicrocircuitAssociative2011, zhangLongrangeLocalCircuits2014}.
Therefore, the VIP-SOM-Pyramidal cell circuit provides a mechanism for context-dependent gating of learning (\textbf{Fig.} \ref{fig:fig_inhibitory_microcircuit}).
When behaviorally relevant signals activate VIP neurons, SOM inhibition of the apical shaft is temporarily lifted, allowing synapses to be modified based on local dendritic computations, which could represent prediction errors or relevant associations \cite{sacramentoDendriticCorticalMicrocircuits2018}.

%%% FIGURE
%%% END FIGURE

\subsubsection{Implementing inhibitory interneurons in ANNs}
It is noteworthy that the power of these microcircuits may lie in the combined integration of global (neuromodulatory) and local error signals.
This also suggests that these two types of error signals may not be disentangled in biological neural networks, while they rarely co-exist in current AI models.
While computational models of these inhibitory interactions have been proposed \cite{wagatsumaMicrocircuitModelInvolving2023}, integrating these concepts in deep ANNs remains an active area of research.
Remarkably, ANNs that respect Dale's principle can be trained with the same efficiency as regular non-Dalean ANNs as long as inhibition centres and standardizes the excitatory signals \cite{cornfordLearningLiveDales2021}.
This principle can also be extended to recurrent neural networks, further supporting the potential for alignment with biological neural networks \cite{liLearningBetterDales2023}.

\subsection{Biologically plausible credit assignment}
Assuming that a useful error signal is available, the next problem is to identify which parameters need to be updated to improve output behaviors.
Credit assignment is a key principle in learning and can be understood at different levels of computations.
The simplest analogy is that of a game of chess, where credit assignment refers to assigning which move(s) contributed most to winning or losing the game.
This concept is also applied directly to units of computations, neurons, at inference time: i.e. which neurons contributed most to successful decisions and adaptive behaviors.
As learning itself involves changes at the synaptic levels though, credit assignment is most useful when considered as identifying which parameters most contribute improving performance.
Credit assignment is one of the most active yet challenging areas of research in machine learning and neuroscience.
As mentioned above, an interaction between feedback and neuromodulatory signals could effectively provide credit assignment in the brain \cite{roelfsema_control_2018}.
Currently however, standard deep learning systems still rely of backpropagating error gradients throughout the entire network, which is famously biologically implausible \cite{rumelhartLearningRepresentationsBackpropagating1986, lecunTheoreticalFrameworkBackpropagation1988}.
While effective in many applications, there are several reasons for which alternatives to backpropagation, as well as establishing clearly what credit assignment mechanisms are used in the brain, could be valuable.

\subsubsection{Backpropagation}
Backpropagation (BP) is efficient at computing gradients and updating weights in multi-layered networks and has many advantages over other learning algorithms in terms of scalability and generalization \cite{rumelhartLearningRepresentationsBackpropagating1986, lecunTheoreticalFrameworkBackpropagation1988}.
However, BP is famously biologically implausible \cite{lillicrapBackpropagationBrain2020a, bengioBiologicallyPlausibleDeep2016, richardsDeepLearningFramework2019}.
This implausibility has been known since its inception and extensively described in the literature, and here I will only briefly summarize the main points.
Firstly, BP requires the computation of exact gradients of the loss function with respect to the weights, which is not how learning is believed to occur in biological systems.
Secondly, BP relies on the assumption of symmetric weights, where the weights used for forward propagation are the same as those used for backward propagation.
One major issue is the "weight transport problem," where BP requires the error signals to be propagated backward through the network using the exact transpose of the weights used in the forward pass \cite{grossbergCompetitiveLearningInteractive1987, lillicrapBackpropagationBrain2020a, liaoHowImportantWeight2016}.
This requirement of precise symmetry between forward and backward connections is not supported by biological evidence, as synapses in the brain are generally unidirectional.
Thirdly, BP requires that error signals at each layer is computed based on the output of the entire network, which is not how learning occurs in biological systems \cite{lillicrapBackpropagationBrain2020a, bengioBiologicallyPlausibleDeep2016, richardsDeepLearningFramework2019}.
Finally, BP explicitly separates training and inference in ways that is unlikely to occur in the brain.
I will cover this final point separately when looking at continual learning.
While BP still remains the state-of-the-art approach to train ANNs, several alternatives have shown promising results.

\subsubsection{Feedback alignment}
One core idea to implement credit assignment without BP is to provide feedback with a dedicated network.
Early attempts include Feedback Alignment (FA) \cite{noklandDirectFeedbackAlignment2016} is one such approach that offers a surprisingly simple yet effective way to circumvent the need for weight symmetry.
The core idea of FA is to use a fixed random matrix for the feedback pathway instead of the transpose of the feedforward weight matrix.
The feedforward weights then learn to adjust themselves such that the error signal propagated through the random feedback matrix provides a useful direction for learning. 
Despite the seemingly arbitrary nature of the feedback weights, FA has shown to be surprisingly effective on various tasks, suggesting that precise weight symmetry might not be essential for learning in deep networks.

\subsubsection{Sign symmetry}
Another another approach is to relax the weight symmetry requirement by using the sign of the feedforward weights for the feedback pathway, rather than the exact values \cite{liaoHowImportantWeight2016}. 
Here the magnitude of the feedback weights does not need to match the forward weights; only the sign is shared. 
This approach is biologically more plausible than strict weight symmetry, as it only requires the direction of influence (excitatory or inhibitory) to be consistent between the forward and backward paths.
Interestingly, networks trained with sign symmetry have also demonstrated promising performance on challenging datasets like ImageNet, approaching the accuracy of backpropagation in some cases. 
The success of feedback alignment and sign symmetry highlights that relaxing the strict weight symmetry constraint of backpropagation while still providing a meaningful error signal can lead to effective learning in deep neural networks.

\subsubsection{Target propagation}
Instead of propagating error gradients backward, Target Propagation (TP) suggests propagating target values for the neural activations \cite{bengioHowAutoEncodersCould2014b, lecunLearningProcessAsymmetric1986, leeDifferenceTargetPropagation2015b, bartunovAssessingScalabilityBiologicallyMotivated2018, ororbiaBiologicallyMotivatedAlgorithms2019b, bengioDerivingDifferentialTarget2020a}.
In this framework, each layer aims to achieve a specific target value, and these targets are determined by propagating information backward from the output layer. 
A common approach in TP involves using layer-wise autoencoders, where each layer learns to reconstruct its input, and the targets for a layer are generated based on the reconstructed activity of the subsequent layer. 
The parameters of the feedforward network are then updated locally to move the actual activations closer to these targets. 
While the concept of target propagation offers a biologically appealing alternative, the initial implementations faced challenges, particularly due to the imperfectness of the learned inverse mappings or autoencoders \cite{leeDifferenceTargetPropagation2015b, meulemansTheoreticalFrameworkTarget2020, bengioDerivingDifferentialTarget2020a}.

\subsubsection{Difference target propagation}
To address these issues and improve the stability and performance of target propagation, Difference Target Propagation (DTP) was introduced \cite{leeDifferenceTargetPropagation2015b}.
DTP incorporates a linear correction mechanism to the target propagation process.
The target for a layer in DTP is not solely based on the output of the feedback network but also takes into account the current activation of that layer and the target of the subsequent layer.
This difference correction helps to stabilize the training process and allows DTP to achieve performance levels that are more comparable to backpropagation. 
The introduction of DTP marked a significant advancement in the field, demonstrating a more robust and effective approach to biologically plausible learning through target-based mechanisms.

\subsubsection{Direct difference target propagation}
Another direction in the evolution of target propagation has led to the exploration of Direct Difference Target Propagation (DDTP; \cite{meulemansTheoreticalFrameworkTarget2020}.
In contrast to the layer-by-layer feedback in standard DTP, DDTP investigates the use of direct feedback connections from the output layer to each of the hidden layers.
This approach aims to streamline the backward propagation of target information and potentially improve the efficiency of the learning process.
Furthermore, advancements in the feedback mechanisms used within target propagation have resulted in the development of specialized loss functions, such as the Local-Difference Reconstruction Loss (L-DRL; \cite{ernoultScalingDifferenceTarget2022}. 
L-DRL is a feedback loss function designed to improve the performance of DTP by facilitating a better alignment between the feedback weights and the effective error signals needed for learning in the feedforward pathway.
DTP, when combined with L-DRL, has achieved remarkable results on image classification tasks, with performance levels that are increasingly comparable to those of backpropagation on the same architectures.
These findings underscore the importance of a feedback mechanisms and loss functions in target propagation to effectively guide learning in a biologically plausible manner.

\subsubsection{Fixed-weight difference target propagation}
Further refinements of Difference Target Propagation have led to the development of Fixed-Weight Difference Target Propagation (FW-DTP) \cite{shibuyaFixedWeightDifferenceTarget2022}.
A key characteristic of FW-DTP is that it keeps the feedback weights constant throughout the training process.
This simplification offers several advantages over standard DTP, where the feedback weights are typically learned alongside the feedforward weights.
By fixing the feedback weights, FW-DTP significantly reduces the computational cost associated with training, as it eliminates the need to update the feedback pathway. 
Despite this simplification, FW-DTP has been shown to still effectively deliver informative target values to the hidden layers of the network, enabling learning to occur.
FW-DTP can achieve improved stability during training and often exhibits higher test performance compared to standard DTP on various image classification datasets.
This suggests that the core concept of target propagation, particularly with the difference correction mechanism, can function effectively even without the need for a learned feedback pathway.
The success of FW-DTP indicates that optimizing the objectives of layer-wise target reconstruction, which often involves training feedback weights, might not be strictly necessary for the concept of target propagation to work properly.

These algorithms based on target propagation highlight the potential role of feedback connections.
It is noteworthy that in the neocortex and throughout most brain regions, neural networks form not only feed-forward, but also feedback (top-down) projections.
Multiple roles have been attributed to these top-down projections, including attention \cite{lindsayAttentionPsychologyNeuroscience2020}, perception \cite{choksiBraininspiredPredictiveCoding2020}, representing predictions \cite{raoPredictiveCodingVisual1999a, raoActivePredictiveCoding2023}
Beyond credit assignment during learning, bidirectional architectures (that take inspiration from the neocortex and include both bottom-up and top-down pathways) have significant benefits in representation learning.
Specifically, top-down signals can be leveraged to resolve ambiguity given that contextual information is available \cite{islahLearningCombineTopdown2023}.
Similarly, context-invariant representations can be learned by feedback rather than feedforward processing \cite{naumannInvariantNeuralSubspaces2022a}.

%%%%% TABLE
\end{multicols}
\begin{table}[H]
\centering
\label{tab:learning_algorithms_compared}
\begin{tabular}{p{0.25\textwidth} p{0.35\textwidth} p{0.3\textwidth}}
\toprule
\textbf{Algorithm} & \textbf{Achievement} & \textbf{Main Limitation} \\
\midrule
Feedback Alignment (FA) & A biologically plausible alternative to backpropagation that uses fixed, random backward weights to align with the forward weights. & Performance can be slightly lower than backpropagation on complex tasks; convergence can be sensitive to network architecture and initialization. \\
\addlinespace
Sign Symmetry & A biologically plausible alternative to backpropagation, leveraging symmetric error signs rather than weight for credit assignment. & Requires careful initialization and training; convergence can be sensitive to hyperparameters. \\
\addlinespace
Target Propagation (TP) & Addresses the non-local credit assignment problem in deep networks by propagating "targets" from the output layer backward. & Can be complex to implement; the "inverse" function required for target propagation might be difficult to learn or define for arbitrary layers. \\
\addlinespace
Difference Target Propagation (DTP) & An extension of TP that focuses on minimizing the difference between the actual and target activations, simplifying the target generation. & Still relies on the concept of inverse operations, which can be challenging to learn accurately, especially in highly non-linear networks. \\
\addlinespace
Direct Difference Target Propagation (DDTP) & A variant of DTP that aims to avoid explicitly learning inverse functions by directly training a "backward" network. & Can still require careful tuning of the backward network; the backward pathway needs to be learned effectively. \\
\addlinespace
Fixed-Weight Target Propagation (FWTP) & Explores target propagation with fixed backward weights (or a fixed backward pathway), simplifying the learning process. & The fixed backward weights might limit the flexibility and representational power compared to learned backward pathways; performance can be sensitive to the choice of fixed weights. \\
\bottomrule
\end{tabular}
\caption{\textbf{Alternatives to backpropagation.} Comparison of algorithms that do not rely on error gradient backpropagation and present biological plausibility.}
\end{table}
\begin{multicols}{2}
\raggedcolumns
%%%%% TABLE

\subsection{Behavioral timescale plasticity and burst-prop}
Assuming that an appropriate error signal can be derived during learning and that credit assignment can be performed appropriately, the next problem is to find an effective approach to update synapses.
Note that in effect, credit assignment and synaptic updates are not completely separable.
In his seminal work \cite{hebbFirstStagePerception1949}, Donald Hebb posited that when two neurons are repeatedly activated in close temporal proximity, the connection between them is strengthened.
This principle, often summarized as "neurons that fire together, wire together," suggested a cellular mechanism by which experiences could induce lasting changes in neural circuits, forming the basis of learning and memory.

\subsubsection{Long-term potentiation}
Experimental evidence supporting Hebbian-like plasticity emerged decades later with the discovery of long-term potentiation (LTP) \cite{blissLonglastingPotentiationSynaptic1973}.
In these experiments, a high-frequency train of stimuli applied to a neural pathway could lead to a long-lasting increase in the strength of synaptic transmission for that pathway.
While LTP demonstrated that synaptic strength could be modified based on activity, it was a relatively coarse measure of correlated firing. 

\subsubsection{Spike-time-dependent plasticity}
The next significant development was the concept of spike-time-dependent plasticity (STDP).
STDP refined the Hebbian principle by demonstrating that the relative timing of pre- and postsynaptic action potentials was critical in determining the direction and magnitude of synaptic plasticity. 
Specifically, if a presynaptic spike consistently precedes a postsynaptic spike within a narrow time window, the synapse is potentiated (LTP).
Conversely, if the postsynaptic spike precedes the presynaptic spike, the synapse is depressed (LTD) \cite{markramNetworkTuftedLayer1997, biSynapticModificationsCultured1998a}.
This timing-dependent rule offered a more nuanced mechanism for how neurons could learn about the temporal relationships between their inputs and their own firing.

One limitation is that the standard STDP model primarily operates on a millisecond timescale, determined by the precise timing of individual spikes.
However, many forms of behavioral learning and memory formation occur over much longer timescales, ranging from seconds to minutes.
It has been questioned whether millisecond-precision STDP alone can fully account for plasticity engaged during complex behaviors \cite{lismanQuestionsSTDPGeneral2010}.
Additionally, the strict requirement for precisely timed pre- and postsynaptic spikes might be less robust in the noisy and asynchronous environment of a living brain.
Finally, these principles have largely been established in the in vitro (brain slices) context and may not translate in vivo as (1) they require a high number of stimulus-response pairings, which are typically not possible in fast-learning contexts, and (2) calcium concentration, and thus neurotransmitter release probabilities typically differ between in vivo and in vitro settings \cite{inglebertCalciumSpikeTimingDependent2021, chindemiCalciumbasedPlasticityModel2022}.

\subsubsection{Behavioral timescale plasticity}
These limitations have paved the way for the investigation of plasticity mechanisms operating on longer timescales, leading to the emergence of the concept of behavioral timescale plasticity (BTSP) \cite{bittnerConjunctiveInputProcessing2015, bittnerBehavioralTimeScale2017}.
BTSP proposes that synaptic plasticity can be driven by events occurring over seconds, a timescale more aligned with natural behaviors and learning processes.
While the precise mechanisms of BTSP are still being elucidated, it is thought to involve the integration of synaptic inputs with longer-lasting postsynaptic depolarizations, such as plateau potentials, which can gate plasticity over extended periods \cite{bittnerConjunctiveInputProcessing2015, bittnerBehavioralTimeScale2017}.
BTSP is particularly interesting for several reasons.
Firstly, its operating timescale aligns well with the temporal dynamics of behavioral learning, providing a more plausible link between neural plasticity and the formation of memories in a naturalistic setting.
Secondly, BTSP in hippocampal neurons has been shown to be crucial for the formation of stable spatial representations (place fields) during navigation, even with single learning trials \cite{bittnerConjunctiveInputProcessing2015, milsteinBidirectionalSynapticPlasticity2021}.
This suggests a mechanism for rapid, one-shot learning that is difficult to explain with standard millisecond-based STDP. 
Finally, recent research indicates that BTSP can also contribute to the formation of non-spatial representations, suggesting a more general role in hippocampal function beyond spatial coding \cite{dorianBehavioralTimescaleSynaptic2025}.
BTSP represents a significant step towards understanding how neural circuits can integrate information over behaviorally relevant timescales to support learning and memory.

\subsubsection{Role of backpropagating associated calcium spikes in plasticity}
It is also important to briefly mention the cellular and physiological principles behind these plasticity events, which will have implications down the line when extrapolating learning algorithms from cellular interactions.
Backpropagating action potentials (BAPs) are electrochemical signals that travel from a neuron's soma back into its dendritic tree \cite{stuartActivePropagationSomatic1994}.
This backpropagation actively depolarizes dendritic branches, though the extent can vary between basal and apical dendrites, with more complex dynamics in the distal apical tuft \cite{gulledgeHeterogeneityPhasicCholinergic2007}.
Calcium ions (Ca2+) are critical for synaptic plasticity and other functions \cite{berridgeNeuronalCalciumSignaling1998}.
A key mechanism linking electrical activity to plasticity is the coincidence detection of synaptic input (EPSPs) and postsynaptic BAPs.
This co-occurrence leads to sufficient depolarization to relieve the magnesium block of NMDA receptors, allowing Ca2+ influx \cite{larkumSynapticIntegrationTuft2009, biSynapticModificationsCultured1998a}.
The precise timing of the EPSP relative to the BAP determines the magnitude and duration of this calcium signal, underpinning plasticity mechanisms such as STDP \cite{markramNetworkTuftedLayer1997}.

In addition to the calcium entry directly associated with BAP depolarization, the conjunction of strong synaptic input and BAPs, particularly in apical dendrites, can trigger regenerative backpropagating associated calcium (BAC) spikes \cite{larkumCalciumElectrogenesisDistal1999} % add (Schiller et al., 1997).
These BAC spikes involve substantial calcium influx through both NMDA receptors and voltage-gated calcium channels, generating a larger and more sustained calcium transient within the dendrite \cite{nevianSpineCa2Signaling2006}.
The amplitude and temporal profile of this intracellular calcium signal act as a crucial switch, determining the specific form of plasticity induced \cite{lismanMechanismHebbAntiHebb1989}.

Ultimately, calcium signals generated by BAPs and BAC spikes translate into changes in synaptic strength by activating different downstream signaling pathways.
High levels of calcium tend to activate protein kinases such as CaMKII, which promote the insertion of AMPA receptors into the postsynaptic membrane, leading to LTP \cite{malenkaLTPLTDEmbarrassment2004a}.
These activity-dependent modifications of synaptic efficacy are fundamental cellular processes that enable neural circuits to adapt and are essential for learning and memory.
In the next section, I will briefly review how these known physiological principles can be translated into learning rules for artificial systems.

\subsubsection{Silence, single, and bursting activity: ternary code of biological neural networks}
Neurons in the brain do not always fire individual spikes; they can also fire in rapid sequences of action potentials known as bursts \cite{lismanBurstsUnitNeural1997}.
Burst firing can induce more significant changes in the strength of synaptic connections compared to isolated single spikes.
This phenomenon, known as burst-dependent plasticity, suggests that bursts play a crucial role in learning and information processing in the brain.
In effect, this suggests that rather than a binary code, neurons could leverage a ternary code of silence, bursts, and single action potentials \cite{naudFastBurstFraction2024}.
The biological significance of neuronal bursts as a mechanism for learning and communication has inspired the development of burst-propagation learning algorithms in ANNs, aiming to leverage these principles for more biologically plausible and potentially more powerful learning methods.

Building upon the biological evidence of burst-dependent plasticity, bursting activity has been proposed as a candidate mechanism to provide temporal credit assignment \cite{gutigSpikingNeuronsCan2016}.
One formalism of this idea is Burstprop, which introduces regularizing feedback connections for each neuron and the use of two types of activity patterns to communicate signed error information \cite{payeurBurstdependentSynapticPlasticity2021}.
This allows Burstprop to be applied to tasks like image classification on the MNIST dataset using networks with hundreds of neurons, a significant step towards handling more realistic problems.
Notably, Burstprop utilizes feedback alignment methods for transporting error signals backward through the network, which is a biologically plausible way to address the weight transport problem.
Experimental results have shown that Burstprop can achieve low test classification errors on MNIST, with performance levels that are comparable to those obtained using backpropagation through time on the same network architectures. 
This demonstrates the potential of burst-dependent learning, as implemented in Burstprop, to scale to more complex learning tasks and serve as a viable method for learning with neuromorphic hardware (which we will briefly cover later).

An intriguing aspect of Burstprop is that it can approximate error backpropagation, effectively performing a similar form of credit assignment but through a more biologically plausible mechanism. 
In this context, neuron bursts can be viewed as acting as teaching signals, conveying information about the error in the network's output, similar to how error signals are propagated in backpropagation. 

\subsection{The continual learning challenge}
Training in deep learning is usually done in batches and separated entirely from inference, where synaptic weights remained fixed.
While for many practical applications, this separation is mostly desirable, it poses problems in the context of continual learning, where learning systems are expected to accumulate knowledge over various tasks and contexts without forgetting previous knowledge.
At the end of training, deep learning models should not only perform well on input close to the training set, but also on never-seen before input, provided that it is part of the underlying distribution learned during training.
Out-of-distribution inputs are usually not well processed in these conditions, and require additional training to extend the learned distribution to new scenarios.
This currently still poses a significant challenge for current neural networks, as re-training a model on new tasks will usually lead to the loss of previous knowledge - a phenomenon commonly referred to as 'catastrophic forgetting' \cite{mccloskeyCatastrophicInterferenceConnectionist1989, ratcliffConnectionistModelsRecognition1990, frenchCatastrophicForgettingConnectionist1999a}.
This section is not meant to cover the continual learning problem extensively, but rather mention relevant brain-inspired approaches. For a more extensive account, see \citet{wangComprehensiveSurveyContinual2024}.

\subsubsection{Continual learning with structural changes}
One set of mitigation approaches focuses on promoting sparse representations, where only a small fraction of neurons are active for any given input, including dropout \cite{hinton_improving_2012} or the k-Winner-Take-All (k-WTA) activation function, which encourage such sparsity \cite{sarfrazSparseCodingDual2023}.
Another approach is to embed some form synaptic stability directly into the model.
In the brain, synapses are situated on dendritic spines.
These spines undergo distinct morphological stages during plasticity: thin filopodial spines are new and unstable, whereas mushroom-shaped spines are stable and less likely to decay \cite{yusteMorphologicalChangesDendritic2001}. 
In deep learning, approaches like Synaptic Intelligence \cite{zenkeContinualLearningSynaptic2017b} or elastic weight consolidation \cite{kirkpatrick_overcoming_2017} aim to protect important synapses from being overwritten when learning new tasks.
Surprisingly, implementing biologically-inspired synaptic turnover mitigates the progressive loss of plasticity observed in connectionist models learning continually \cite{dohare_loss_2024}.

\subsubsection{Continual learning with replay}
\begin{figure}[H]
  \includegraphics[width=\columnwidth]{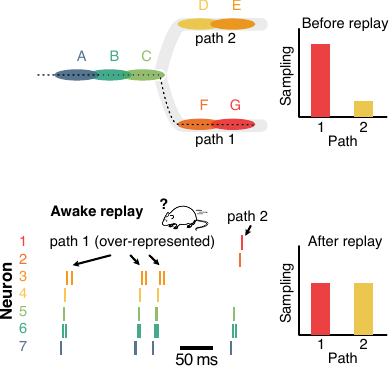}
  \captionsetup{width=\columnwidth}
  \caption{
    \textbf{Role of replay in continual learning}.
    Neuronal activities associated with awake exploration are replayed during periods of inactivity (quiet wakefulness).
    While replay has often been associated with memory consolidation (during sleep), an alternative hypothesis is that representing under-sampled experiences in an offline state could correct for biases in the 'training data' of sampled explorations \cite{etterLinkingTemporalCoordination2023a, meerAwakeReplayClock2025}.
}
  \label{fig:fig_replay_bias}
\end{figure}

Another important class of solutions for continual learning takes direct inspiration from replay.
While animals are believed to learn continuously, this does not mean that training and inference completely overlaps, as mammals go through sleep-wake cycles that are associated with distinct brain states, including REM and slow-wave sleep.
In animals, including humans, neurons tend to replay awake patterns during sleep \cite{skaggsReplayNeuronalFiring1996, leeMemorySequentialExperience2002, nadasdyReplayTimeCompression1999}.
While replay has traditionally been associated with memory consolidation \cite{jadhavAwakeHippocampalSharpwave2012, girardeauSelectiveSuppressionHippocampal2009, gridchynAssemblySpecificDisruptionHippocampal2020, boyceCausalEvidenceRole2016a}, its exact physiological role remains unknown.
Additionally, replay occurs during wakefulness, where it could have roles beyond consolidation.
One recent hypothesis is that of awake replay improving future goal-oriented behavior by internally simulating past experiences, then using them to adjust estimates of future rewards, effectively biasing neural activities towards prospective rewards \cite{meerAwakeReplayClock2025} (\textbf{Fig.} \ref{fig:fig_replay_bias}).
Congruent with this hypothesis has been the application experience replay to encourage ANNs to maintain similar predictions for the replayed samples \cite{sarfrazSparseCodingDual2023}.
While most experience replay approaches either directly memorize previous samples \cite{chaudhry_tiny_2019, riemer_learning_2019, lopez-paz_gradient_2017, rebuffi_icarl_2017, aljundi_gradient_2019, borsos_coresets_2020, yoon_online_2022, shim_online_2021, tiwari_gcr_2022, chaudhry_efficient_2019} or generate previous samples in input space \cite{shin_continual_2017, ostapenko_learning_2019, wang_triple-memory_2022, kemker_fearnet_2018, xiang_incremental_2019, cong_gan_2020, liu_generative_2020}, it is also possible to generatively replay previous samples in latent space \cite{van_de_ven_brain-inspired_2020}, which is more biologically plausible.

%%% FIGURE
%%% END FIGURE

\subsubsection{Continual learning in real-time}
\begin{figure}[H]
  \includegraphics[width=\columnwidth]{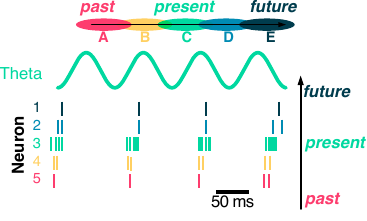}
  \caption{
    \textbf{Predicting the future 8 times per second}.
    As animals explore environments, neurons respond to stimuli (locations, objects) with prominent bursts of action potentials.
    Importantly, neurons representing past and future explorations still fire single action potential in a sequence than spans the spatiotemporal continuum, all within a single theta cycle of a few milliseconds.
    This sequential activity reflects the maintenance of past-present-future representations quasi-simultaneously, and could be relevant for learning new information without forgetting immediate past experiences \cite{etterLinkingTemporalCoordination2023a}.
}
  \label{fig:fig_precession}
\end{figure}

This section will explore more speculative roles of known physiological mechanisms in continual learning.
In vivo, the mammalian brain expresses a wide range of oscillations that reflect the coordinated activity of excitatory and inhibitory neurons \cite{buzsakiMechanismsGammaOscillations2012, wulffHippocampalThetaRhythm2009}.
Importantly, the excitability of individual neurons is modulated by these broader oscillation signals.
Theta oscillations (6 - 12 Hz) in particular may provide temporal windows during which plasticity can occur \cite{etterLinkingTemporalCoordination2023a}.
During awake exploration, neurons representing locations (or objects) tend to do so in sequences that span the past, present, and potential futures.
Neurons representing present exploration tend to burst, whereas neurons representing past and future experiences fire single action potentials (\textbf{Fig.} \ref{fig:fig_precession}).
These so called theta-sequences could introduce robustness against catastrophic forgetting, as both immediate memory and future planning can co-exist quasi-simultaneously during active cognition \cite{etterLinkingTemporalCoordination2023a}.

%%% FIGURE
%%% END FIGURE

When considered together with the aforementioned burst-based learning rules and inhibitory interneuron microcircuits, local field and dendritic oscillations could provide a direct mean for testing future predictions, learning relevant information, while maintaining previous immediate experiences, all in real-time.

\begin{figure}[H]
  \includegraphics[width=\columnwidth]{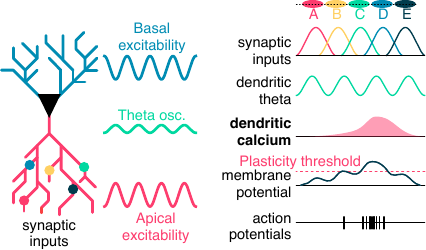}
  \captionsetup{width=\columnwidth}
  \caption{
    \textbf{Real-time calcium-based plasticity}.
    In addition to the interneuron microcircuit discussed earlier, excitatory neurons fire with respect to underlying oscillations.
    Remarkably, these oscillations are differentially expressed in dendrites, and can contribute to the exact timing of output action potentials.
    In this model, I propose that past, current experiences, and future decisions are all encoded in distinct synaptic inputs and expressed sequentially as sub-threshold activity in dendrites.
    Only when future predictions are being test at present time, can dendritic calcium increase to levels that enable bursts of action potentials, synaptic plasticity, and learning \cite{etterLinkingTemporalCoordination2023a}.
}
  \label{fig:fig_calcium}
\end{figure}

The main idea behind this tenet is that distinct basal and apical dendritic compartments oscillate within the theta frequency band, but at distinct oscillatory phases.
Calcium can accumulate specifically in apical dendrites \cite{yusteCa2AccumulationsDendrites1994}, and concomitant basal and apical activations can drive the generation of plateau potentials reliably \cite{larkumNewCellularMechanism1999a, naudSilencesSpikesBursts2023}.
This suggests that in addition to an upstream teaching signal that results from the interaction between interneurons, neuromodulators, and plateau potentials, there are specific temporal windows during which new information can be encoded, as determined by the underlying dendritic oscillations.
Given that theta sequences contain information about immediate past experiences and future goals, and input representations co-exist in dendrites \cite{sheffieldCalciumTransientPrevalence2015, sheffieldDendriticMechanismsHippocampal2019}, this could provide a transient mechanism to support continual learning in real time (\textbf{Fig.} \ref{fig:fig_calcium}).

%%% FIGURE
%%% END FIGURE

\section*{Neuromorphic computing}
The premise of neuromorphic computing is that the brain's architecture and processing principles represent a highly efficient and effective blueprint for certain types of computation, particularly those involving pattern recognition, learning, and adaptation in complex, noisy environments \cite{meadNeuromorphicElectronicSystems1990}.
Current AI models, predominantly based on deep learning running on conventional von Neumann architectures (with separate processing and memory units), has achieved remarkable success but faces significant hurdles.
These include immense energy consumption, the need for vast datasets for training, and limitations in continuous learning and robustness to novel situations.
Neuromorphics proposes a paradigm shift, aiming to build hardware that directly mimics neural structures and dynamics, potentially overcoming these limitations by leveraging the brain's inherent parallelism, event-driven processing, and co-location of memory and computation \cite{schumanSurveyNeuromorphicComputing2017}.

Memristors, or resistive memory devices, have garnered significant attention as potential building blocks for neuromorphic synapses due to their ability to change resistance based on the history of voltage applied or current passed through them, and retain that state \cite{chuaMemristorTheMissingCircuit1971, strukovMissingMemristorFound2008}.
Their key advantages include nanoscale size (allowing for extremely dense memory arrays), analog (multi-level) resistance states (ideal for representing synaptic weights), non-volatility (retaining memory without power), and the potential for integration into crossbar arrays that naturally perform vector-matrix multiplications crucial for neural networks, thus enabling efficient in-memory computing and mitigating the von Neumann bottleneck \cite{ielminiInmemoryComputingResistive2018}. 
Basic STDP-like behavior, pattern classification, and other useful synaptic functions have been successfully implemented using memristor arrays \cite{joNanoscaleMemristorDevice2010, preziosoTrainingOperationIntegrated2015}.

Despite the promise, significant challenges remain.
Memristive devices often suffer from considerable variability (device-to-device and cycle-to-cycle fluctuations in behavior), limited endurance (number of times they can be reliably switched), non-ideal switching characteristics (non-linearity, asymmetry between potentiation and depression), and difficulties in achieving precise, incremental weight updates needed for many learning algorithms \cite{markovicPhysicsNeuromorphicComputing2020, schumanSurveyNeuromorphicComputing2017}.
Integrating these nanoscale devices reliably with standard CMOS circuits for control and readout also poses manufacturing challenges.
Overcoming these hurdles is critical for realizing large-scale, reliable memristor-based neuromorphic systems.

One of the most compelling arguments for neuromorphic computing is its potential for energy efficiency.
Training state-of-the-art deep learning models, especially massive language models, requires enormous computational resources, consuming megawatts of power for weeks or months, resulting in substantial carbon footprints  \cite{strubellEnergyPolicyConsiderations2020, pattersonCarbonEmissionsLarge2021}).
The operational (inference) costs are also significant, particularly as AI becomes more pervasive.
This energy demand poses a serious challenge to sustainable AI development and limits the deployment of complex AI on power-constrained devices (e.g. mobile phones, wearables, remote sensors).
Neuromorphic systems, through their event-driven operation (computation only where needed), inherent parallelism, sparse activity, and co-location of memory and processing (reducing costly data movement), promise energy savings potentially orders of magnitude greater than conventional hardware for equivalent tasks \cite{daviesLoihiNeuromorphicManycore2018}.
Achieving this efficiency could reduce the environmental impact of data centers and enable entirely new applications requiring continuous, low-power intelligent processing.

\section*{Conclusions}
Advances in deep learning have been in large part enabled by breakthroughs in neuroscience.
Conversely, deep learning frameworks have provided neuroscientists with computational frameworks to test fundamental questions about how the brain might operate.
Beyond well established neuroscience principles in deep learning, such as nonlinearities and architectures, many recent neuroscience discoveries must be brought to the attention of the machine learning community.
In particular, recent discoveries related to behavioral timestamp plasticity, inhibitory interneuron microcircuits, and real-time in vivo computations with oscillations could provide testing grounds for the next generation of AI systems.

Critically, von Neumann hardware will eventually be limited in its ability to implement biologically plausible artificial systems, in particular those that rely on spike-based learning rules.
Ultimately, progress for these bioplausible algorithms is limited by the very knowledge frontier in neuroscience, and neurophysiological experimentation still holds significant value for AI practitioners.
While still an emerging field, neuromorphics present a valuable potential for running biological computations while circumventing the limitations of von Neumann hardware all the while reducing energy demand, thus tackling some of the biggest challenges ahead of machine learning.

\printbibliography
\end{multicols}
\end{document}